\documentstyle[11pt,cs11,html,epsf]{article}

\begin{document}

\title{
Young stellar kinematic groups
and their relation with young open clusters, star forming regions and \\ 
the Gould Belt
}

\author{D. Montes 
}
\affil{Departamento de Astrof\'{\i}sica, 
Univ. Complutense de Madrid, Spain}

% Notice that some of these authors have alternate affiliations, which
% are identified by the \altaffilmark after each name.  The actual alternate
% affiliation information is typeset in footnotes at the bottom of the
% first page, and the text itself is specified in \altaffiltext commands.
% There is a separate \altaffiltext for each alternate affiliation
% indicated above.

% The nice thing about this method is that it saves space on the first page!

%\altaffiltext{1}{Guest observer at McDonald Observatory}

% BUT if you've used \altaffiltext and you think you'll be adding 
% *footnotes*, then you need to update the footnote counter!

%\setcounter{footnote}{3}

% The abstract is entered in a LaTeX "environment", designated with paired
% \begin{abstract} -- \end{abstract} commands.  Other environments are
% identified by the name in the curly braces.

\begin{abstract}

Stellar kinematic groups (SKG) are kinematically coherent groups of stars
that share a common origin.
We have compiled (Montes et al. 1999; 2000)
a sample of late-type stars of previously established members
and possible new candidates to different young SKG
(Local Association (20 - 150 Myr),
Ursa Mayor group (300 Myr),
 Hyades supercluster (600 Myr),
 IC 2391 supercluster (35 Myr) and
Castor Moving Group (200 Myr)).
In order to better understand the origin of these young SKG, and
to be able to identify late-type stars members of the classical and
the recently identified SKG,
we also need to study the kinematic properties of nearby
young open clusters and star forming regions.
With this aim we have taken the most recent
data available in the literature (including astrometric data from
Hipparcos Catalogue) of the nearby young
open clusters, OB associations, T associations, and
other associations of young stars as TW Hya.
We use these data
to calculate their Galactic space motions (U, V, W)
and space coordinates (X, Y, Z) and study their possible association
with the different SKG as well as with
the young flattened and inclined Galactic structure known as the Gould Belt.

\end{abstract}

% Keywords should be included, but they are not printed in the hardcopy.
% They will be used by the Editors to help organize poster papers by
% category though!


% That's it for the front matter.  On to the main body of the paper.
% We'll only put in tutorial remarks at the beginning of each section
% so you can see entire sections together.

% 
% OK - to make things easier for the Editors, we're going to put
% all of our object aliases up front since we only have to declare
% them once in the paper.  Some people prefer to use NGC 7078 for
% M 15, but we like good old Messier, so that's what we'll index by.
% But we'll cross-reference it here so that people who do like NGC 7078
% won't have to remember that it's also M 15!
%
% Remember - we identify objects by putting an asterisk in front of the name!
%
%\index{*BF Lyn|HD 80715}

\section{Introduction}

It has long been known that in the solar vicinity there are several kinematic 
groups of stars that share the same space motions that well know open clusters.
These stellar kinematic groups (SKG), called moving groups (MG) or 
superclusters (SC), are kinematically coherent groups of stars
that share a common origin (the evaporation of a open cluster, the remnants 
of a star formation region or a juxtaposition of several little star formation 
bursts at different epochs in adjacent cells of the velocity field).
The youngest and best documented SKG are: 
the Hyades supercluster (600 Myr, Eggen 1992b)
the Ursa Mayor group (Sirius supercluster) 
(300 Myr, Eggen 1992a, 1998; Soderblom \& Mayor 1993),
the Local Association or Pleiades moving group
(20 to 150 Myr, Eggen 1992c),
the IC 2391 supercluster (35-55 Myr, Eggen 1995), and
the Castor moving group (200 Myr, Barrado y Navascu\'es 1998).
Since Olin Eggen introduced the concept of moving group and the idea that 
stars can maintain a kinematic signature over long periods of time, their
existence has been rather controversial.
However, recent studies (Dehnen 1998; Chereul et al. 1999; Asiain et al. 1999; 
Skuljan et al. 1999) using astrometric data taken from Hipparcos and different 
procedures to detect MG not only confirm the existence of classical young MG,
but also detect finer structures in space velocity and age.
Well known members to these moving groups are mainly early type stars
and few studies have been centered in late-type stars.
We have compiled  (Montes et al. 1999, 2000) a sample of
late-type stars (single and spectroscopic binaries) of previously established 
members and possible new candidates to these five young SKG. 

%----------------------------------
\begin{figure}
\plottwo{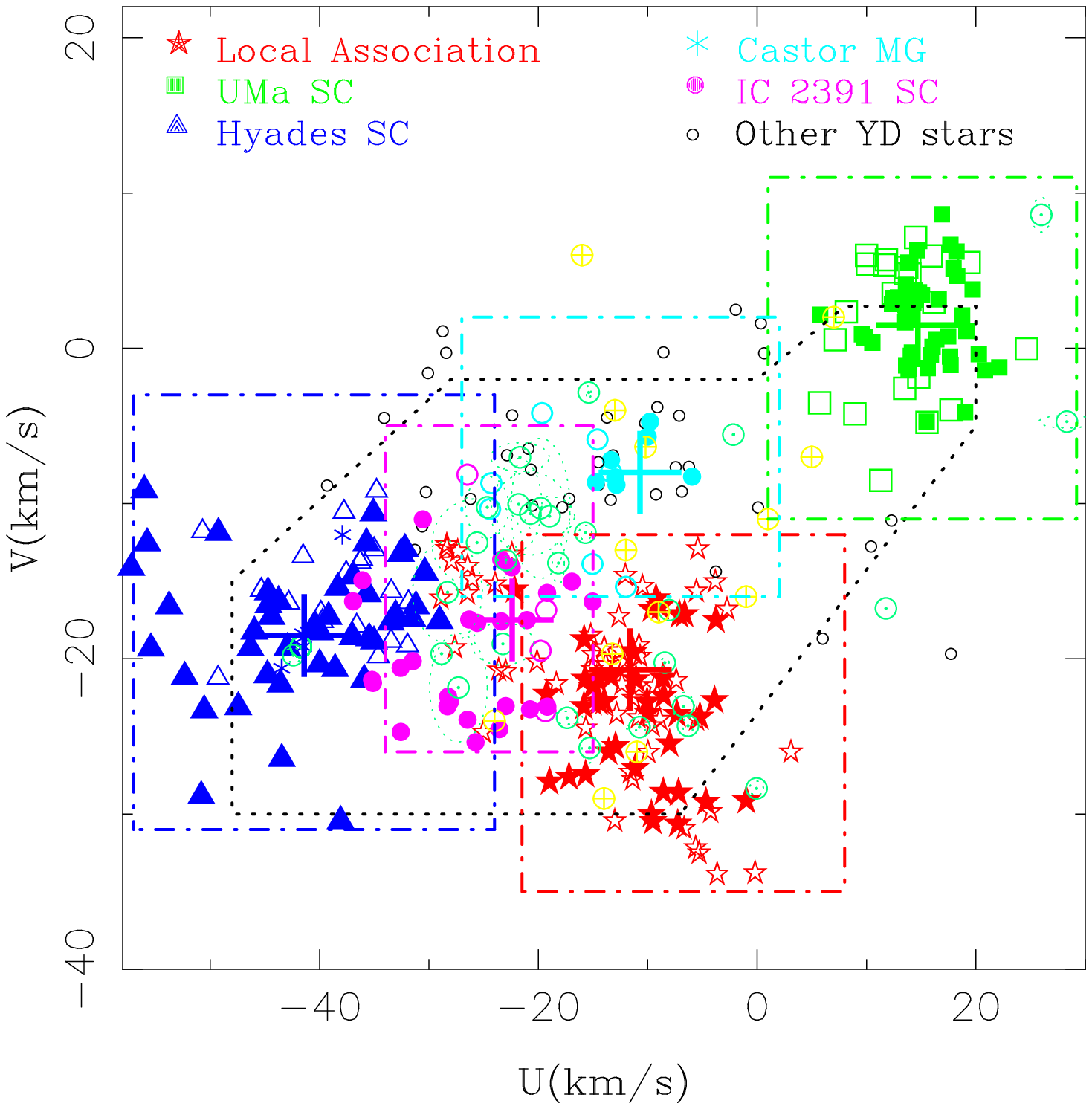}{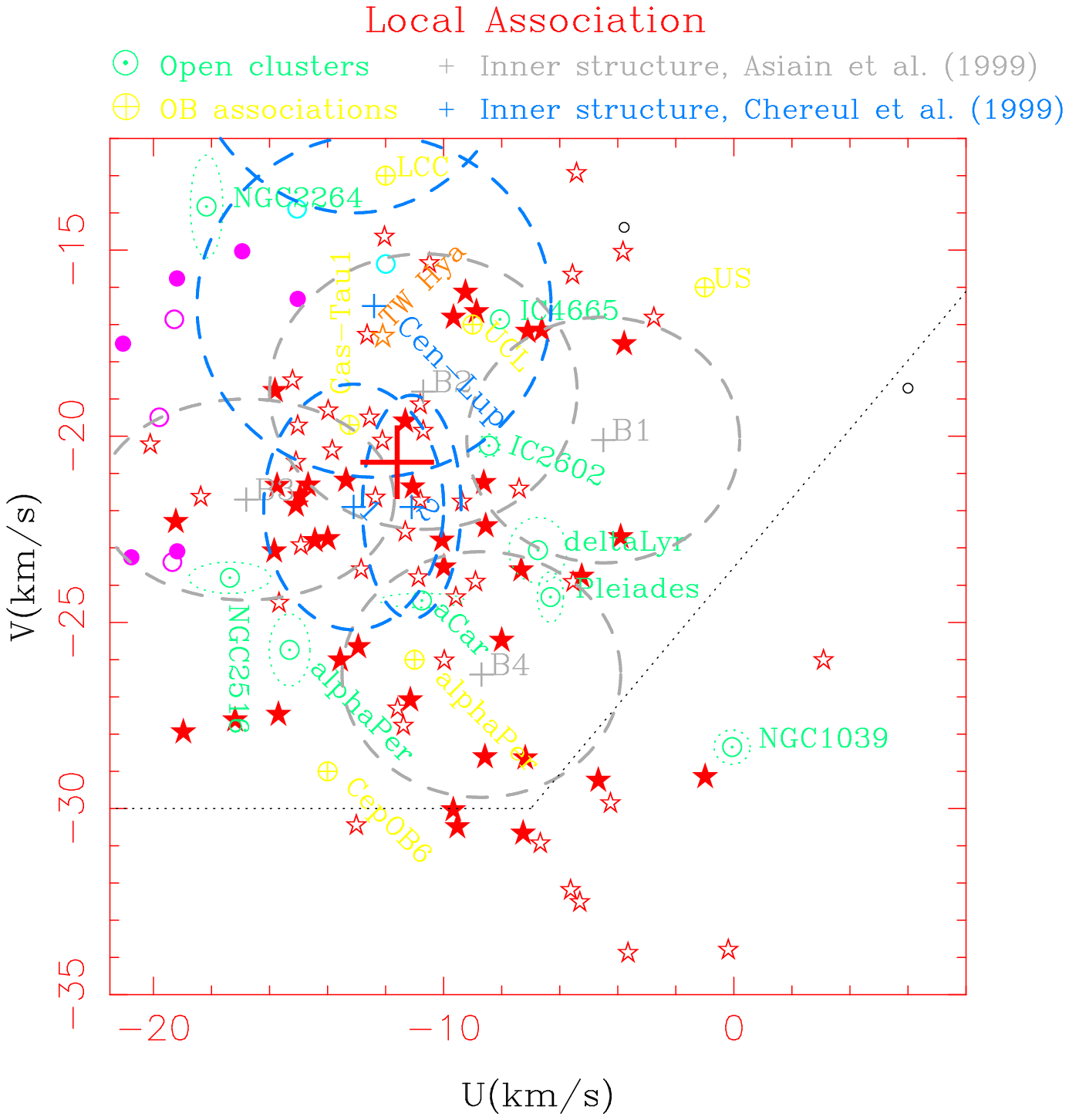}
\caption{
In the left panel we plot the (U, V) diagram for a sample of late-type stars
 (Montes et al. 1999, 2000)
identified as members (filled symbols)  or possible members (open symbols) of
different young stellar kinematic groups (represented by different symbols 
and colors). Big crosses are plotted in the central position of each group.
The black dashed line represents the  boundaries
that determine the young disk population as defined by Eggen (1984, 1989).
The color dotted-dashed rectangles represent the regions enlarged in Fig. 1, 
to 3. The positions of nearby young open clusters ($\bigodot$) and OB 
associations ($\bigoplus$) are also plotted in this diagram.
The right panel is an enlargement of the (U, V) diagram 
in the region of the Local Association.
}
 \label{fig-1}
\end{figure}
%----------------------------------
In order to better understand the origin of these young MG, and
to be able to identify late-type stars members of the classical and
the recently identified MG and substructures,
one also need to study the kinematic properties of nearby
young open clusters and star forming regions.
With this aim I have calculated the galactic space-velocity components 
(U, V, W), using the most recent
data available in the literature (including astrometric data from
Hipparcos Catalogue), of the nearby young
open clusters (Robichon et al. 1999),
OB associations (de Zeeuw et al. 1999), T associations, and
other associations of young stars as TW Hya (Webb et al. 1999).
The position of these different young structures in the UV and WV diagrams is
compared with the position of the classical MG as well as with the position and
associated velocity dispersion of the new substructures recently found by
Chereul et al. (1999) and Asiain et al. (1999).
Finally, the possible relation of all these young stars concentrations
with the young flattened and inclined Galactic structure known as the 
Gould Belt has also been analysed.

%----------------------------------
\section{Young Open Clusters and OB associations}
%----------------------------------

Mean astrometric parameters of nearby young open clusters (d$<$ 500pc) 
have been taken  from  Robichon et al. (1999),
 from Perryman (1998) for the Hyades, Scholz et al. (1999) for IC 348
and Platais et al. (1998) for a Car.
These mean parallaxes and proper motion
have been computed using Hipparcos data (ESA, 1997).
For other young open clusters with no Hipparcos data parameters have been
taken for differences sources (Palou\v{s} et al. 1977).
Other clusters
(as $\delta$ Lyr (Eggen 1968, 1972, 1983) and NGC1039 (Eggen 1983a))
cited in the literature as probably associated with some SKG
have also  been included here.

Table 1.\footnote{Table 1 available at 
{\tt http://www.ucm.es/info/Astrof/oclusters$\_$uvw$\_$tab.html}}
list all the open clusters included in this study in order of increasing age.
I give the name, age (Myr), distance (pc), metallicity, [Fe/H], 
coordinates (FK5 1950.0), and the U, V, W, calculated components with their 
associated errors in km/s.
Age, distance,  and metallicity have been taken from
different sources as given at the end of the table.
 Galactic space-velocity components (U, V , W)
in a right-handed coordinated system (positive in the directions of the 
Galactic center, Galactic rotation, and the North Galactic Pole, respectively) 
and their associated errors have been calculated using the
procedures given by Johnson \& Soderblom (1987).

For nearby OB associations we adopt the space velocities (U, V, W) calculated
by de Zeeuw et al. (1999) using mean astrometric parameters from Hipparcos (ESA, 1997).

%----------------------------------
\begin{figure}
\plottwo{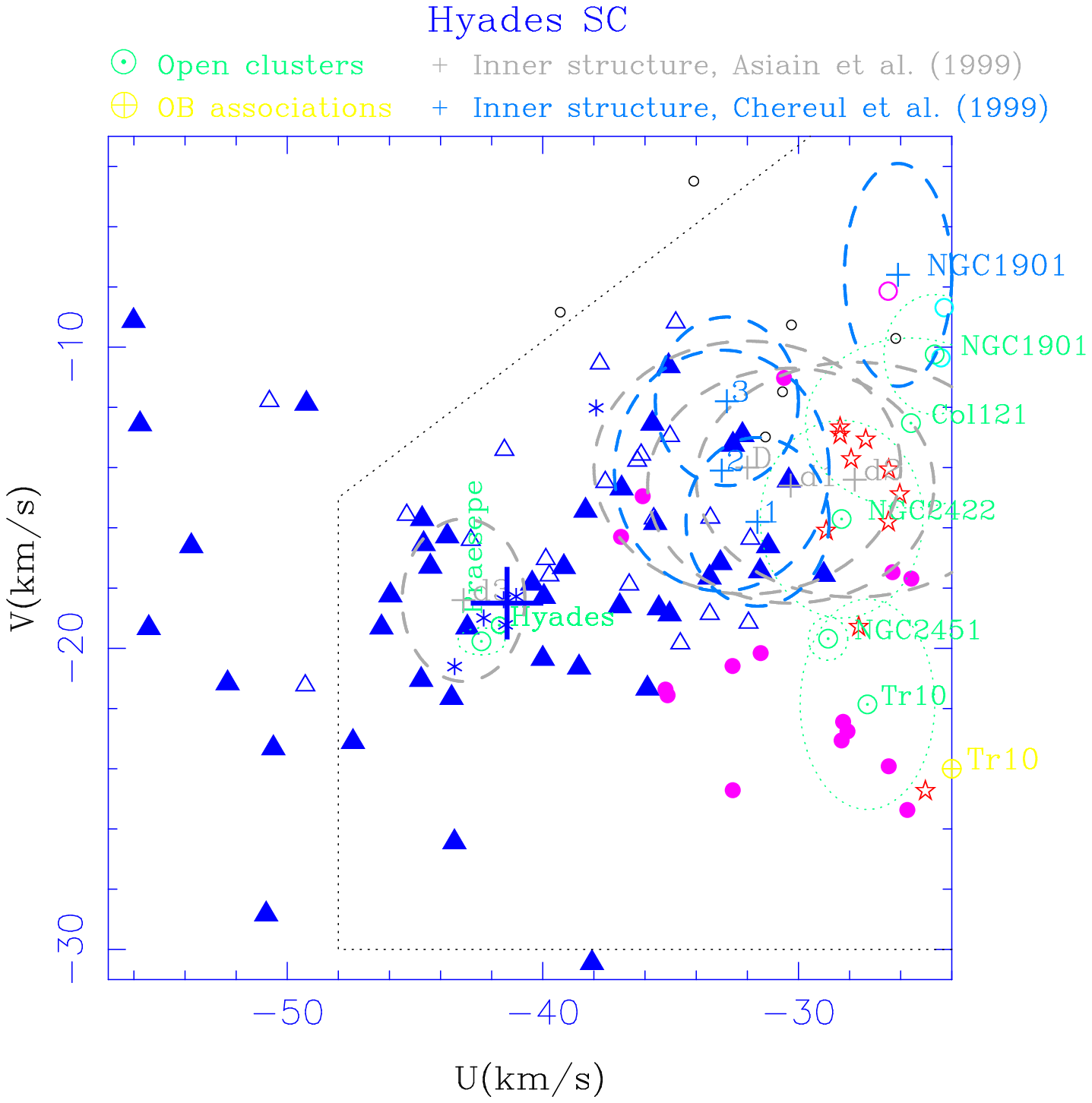}{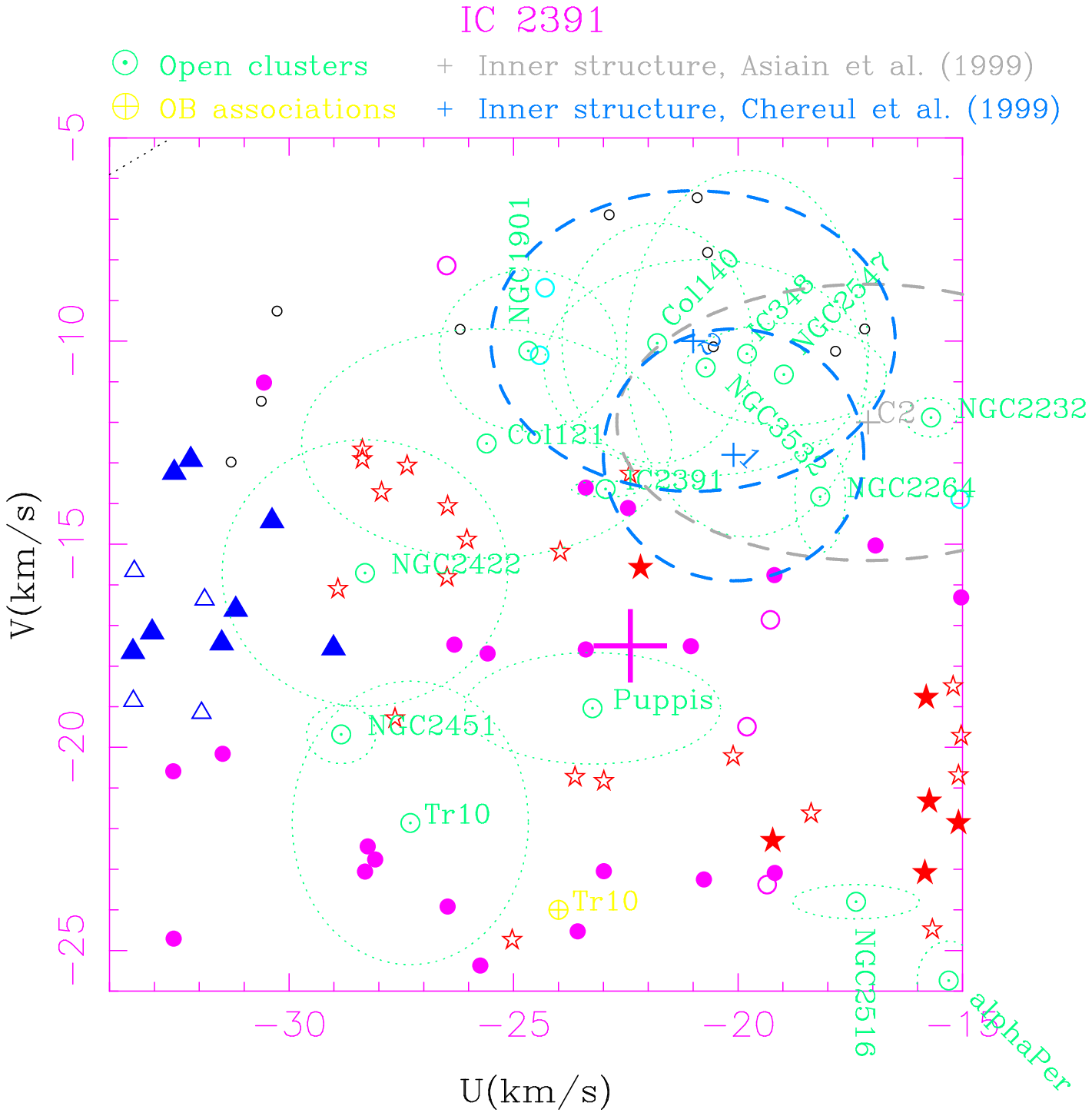}
\caption{
Enlargement of (U, V) diagram in the region of Hyades supercluster (left panel)
and IC2391 supercluster (right panel).
}
 \label{fig-2}
\end{figure}
%----------------------------------

%----------------------------------
\section{Young stellar kinematic groups}
%----------------------------------

The (U, V) velocity components of the five young stellar kinematic groups 
above mentioned are plotted in Fig. 1 (left panel).
All these groups fall inside the boundaries (black dashed line in Fig. 1)
that determine the young disk population as defined by Eggen (1984, 1989).
The different inner structures found by 
Chereul et al. (1999, C99 hereafter) and  
Asiain et al. (1999, A99 hereafter) in these MG, 
as well as possible new MG identified by 
these authors are plotted in Fig. 1 to 3.
The dashed ellipse plotted for each structure represents the associated 
velocity dispersion.
The position of the clusters and OB associations in the UV diagram is plotted 
in Fig. 1 (left panel).
The dotted ellipse plotted around each open cluster position represents the
associated errors in U and V.
Each cluster and OB associations is identified with its name 
in figures 1 to 3.

\underline{\sc  Local Association}
\newline
Several clusters and associations have been suggested as members of the 
Local Association (LA):
The southern concentrations NGC 2516, IC 2602, Upper Centaurus Lupus (UCL) 
and Upper Scorpius (US) (Eggen 1983a) and
the northern concentrations $\alpha$ Per, Pleiades, NGC 1039, 
$\delta$ Lyr (Eggen 1983b).
In Fig. 1 (right panel) we can see that the U, V velocities of all these 
open clusters 
are close to the U, V velocities of the LA, except NGC 1039 whose velocities 
fall even outside of the young disk boundaries.
Other clusters that can also be associted to this MG are IC 4665 and a Car.
The OB association UCL and Cas-Tau 1, and the structure associated to Centaurus 
 Lupus by C99 have velocity components close to the LA, but,
US and Cep OB6  has  U, V velocities a bit different.
A99 found four substructures (B1, B2, B3 and B4) of different 
ages associated to the LA.
As it can be seen in  Fig. 1 (right panel) 
these structures fall well around the central position of the LA.
B1 and B2 are the youngest structures and seems to be related with IC 2602, 
IC 4665, UCL, Cas-Tau 1 and TW Hya.
B4 is associated to the Pleiades cluster, and the oldest structure B4 could be 
related with NGC2516.
The two structures (1 and 2) found by  C99 have also velocity components close 
to the LA.

%----------------------------------
\begin{figure}
\plottwo{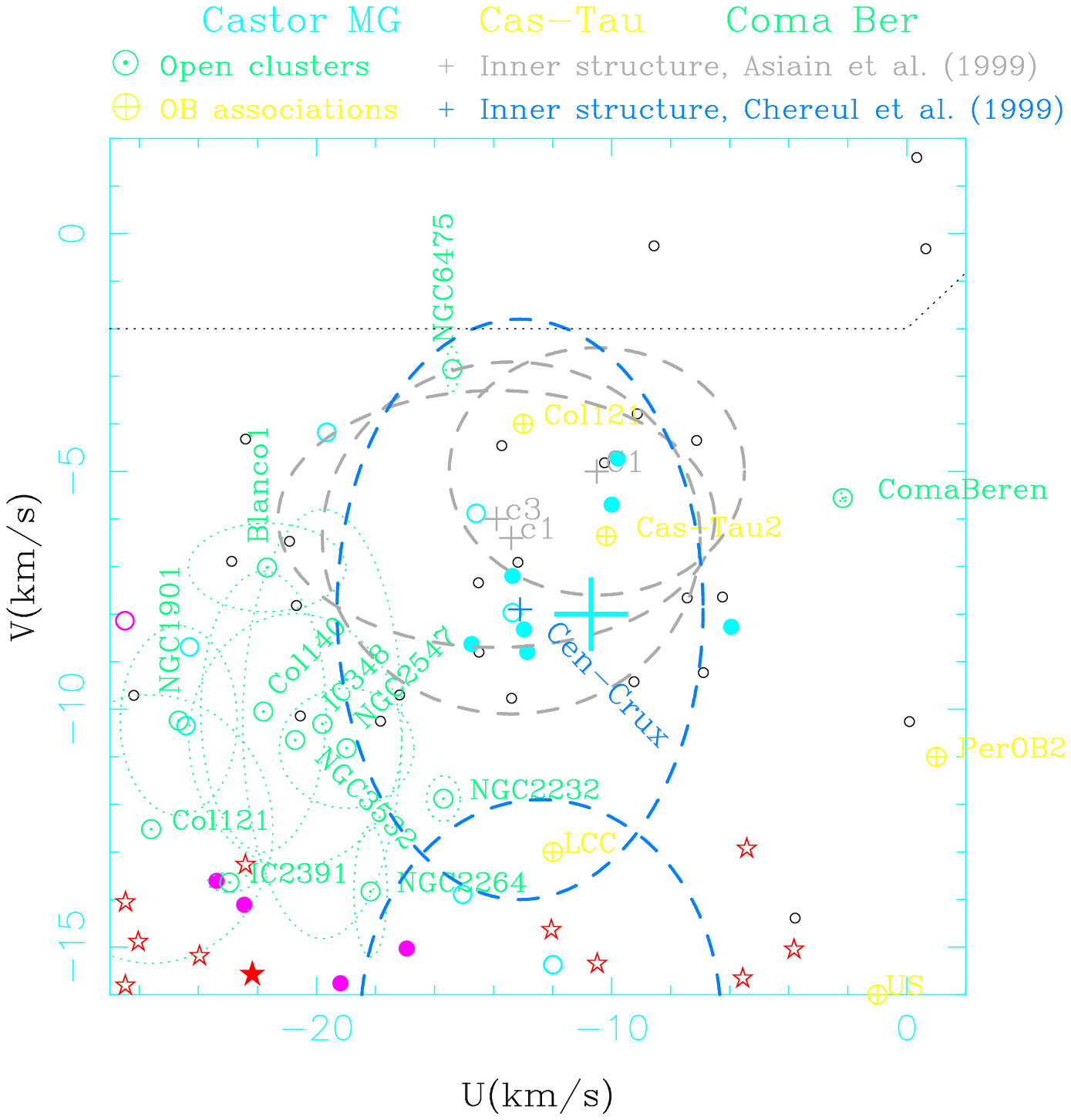}{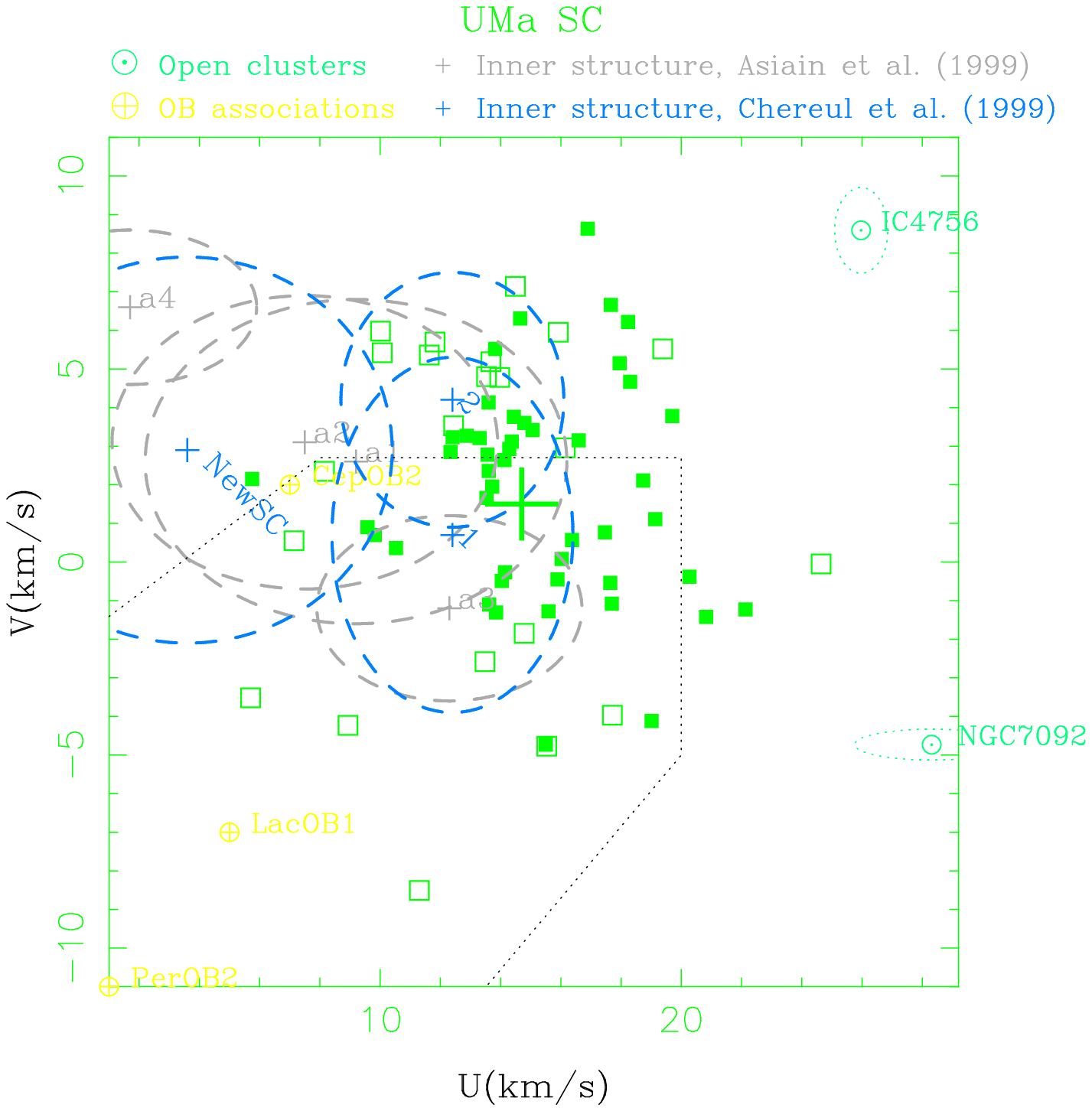}
\caption{
Enlargement of (U, V) diagram in the region of Castor MG (Cast-Tau, Coma Ber)
and Ursa Mayor group (right panel).
}
 \label{fig-3}
\end{figure}
%----------------------------------

\underline{\sc IC 2391}
\newline
In Fig. 2 (right panel) it can be seen that the velocity components 
of the IC 2391 
open cluster as well as the substructures associated by C99 
and A99 to the IC 2391 supercluster are above the (U, V) 
position given for this supercluster. Close in velocity space to IC 2391 open 
cluster and the substructures there are (see Fig. 2 right panel)
 a large concentration of 
open clusters: NGC 2264, NGC 2232, NGC 2547, IC 348, NGC 2532, Col 140, 
NGC 1901, Col 121. Other cluster as NGC 2422, NGC 2451 and Tr 10 
(and Tr 10 OB association) are not in the mentioned concentration but could be 
related with IC 2391 supercluster.

\underline{\sc Castor moving group}
%Castor
%------
\newline
Fig. 3 (left panel) is centered in the (U, V) velocities of 
the Castor moving group.
The structure associated to Centaurus Crux by C99 have 
a very large velocity dispersion and fall in this region together with the 
OB associations Lower Centaurus Crux (LCC), Cas-Tau 2, and Col 121.
The velocity components of the groups c1, c3, and C1 found by 
A99 are more close to Cas-Tau 2 association than to the 
Coma Berenices open cluster (see Fig. 3 left panel).

\underline{\sc Ursa Mayor group}
%UMa
%---
\newline
The open cluster NGC 7091 was suggested by Eggen as possible related with 
the Ursa Mayor group, but as it can bee seen in Fig. 3 (right panel) its (U, V) 
components are very far from the center of the group.
Other cluster not related with the group is IC 4756.
Four substructures (a1, a2, a3, a4) have been found by A99 
in this region.
a1 y a2 have U component lower than the Ursa Mayor group and are close to 
the OB association Cep OB2.
a34 have a very low U component and is more close (see Fig. 3 right panel) 
to the new supercluster identified by C99.
The substructures 1 and 2 of C99 have (U, V) components 
very close to the Ursa Mayor group.

\underline{\sc Hyades supercluster}
%Hyades
%------ 
\newline
The Hyades and Praesepe open clusters have velocity components very very 
close to the (U, V) components of the Hyades supercluster 
(see Fig. 2 left panel).
Eggen (1996) assumed the NGC 1901 cluster as a component of the Hyades 
supercluster, however as it can be seen
in Fig. 2 (left panel) the (U, V) components of the open cluster and of 
the substructure 
associated to NGC 1901 by C99 are very different from the 
supercluster velocity components.
At (U, V) velocities intermediate between the supercluster and NGC 1901 is 
where  C99 found substructures Hyades 1, 2 and 3 and 
A99 found the substructures d1, d2 and D.

%----------------------------------
\section{The Gould Belt}
%----------------------------------

The Gould Belt (Gould 1879; P\"oppel 1997) is a ring-like, expanding, 
flattened and inclined 
Galactic structure outlined by OB associations, molecular gas, 
star forming regions and young galactic clusters.
Recently, Guillout et al. (1998a, 1998b) discovered a galactic latitude 
enhancement of X-ray active stars consistent with the Gould Belt.
Detailed analysis of the surface density, distance and X-ray luminosity 
distributions led these authors to conclude that these stars are  
distributed in a disk disrupted near the Sun (Gould disk).
Wichmann et al. (1997) have also found that the spatial distribution of the
 ROSAT discovered WTTS stars near Lupus, 
perpendicular to the galactic plane, is centered on the Gould Belt.
Many of the young stars concentrations analysed here are associated with the
Gould Belt and therefore this galactic structure could also be the origin
of the young moving groups 
(for a detailed analysis see Montes (2000, in preparation).

%----------------------------------

% Finally, we have a little acknowledgements section.

\acknowledgments

This work has been supported by the Universidad Complutense de Madrid
and the Spanish Direcci\'{o}n General de  Ense\~{n}anza Superior e
Investigaci\'{o}n Cient\'{\i}fica (DGESIC) under grant PB97-0259.

\end{document}